\documentclass[12pt]{article}

\usepackage{scicite}
\usepackage{color}
\usepackage{hyperref}
\hypersetup{
    colorlinks=true,
    linkcolor={blue},
    citecolor={blue},
    urlcolor={blue}
}

\usepackage{ulem}
\usepackage{bm}
\usepackage{authblk}

\usepackage{graphicx}
\graphicspath{ {./images/} }

\usepackage[figurename=Fig.]{caption}

\usepackage{times}

\usepackage{xr}
\makeatletter

\newcommand*{\addFileDependency}[1]{
\typeout{(#1)}
\@addtofilelist{#1}

\IfFileExists{#1}{}{\typeout{No file #1.}}
}\makeatother

\newcommand*{\myexternaldocument}[1]{%
\externaldocument{#1}%
\addFileDependency{#1.tex}%
\addFileDependency{#1.aux}%
}

\myexternaldocument{sm}

\newenvironment{sciabstract}{%
\begin{quote} \bf}
{\end{quote}}

\usepackage{amsmath,amssymb}

\let\originallesssim\lesssim
\let\originalgtrsim\gtrsim

\DeclareRobustCommand{\lesssim}{%
  \mathrel{\mathpalette\lowersim\originallesssim}%
}
\DeclareRobustCommand{\gtrsim}{%
  \mathrel{\mathpalette\lowersim\originalgtrsim}%
}

\def\Rutgersphysics{Department of Physics \& Astronomy, \protect\\
Rutgers University, Piscataway, New Jersey 08854, USA}

\def\Rutgerscmt{Center for Materials Theory, \protect\\
Rutgers University, Piscataway, New Jersey 08854, USA}

\def\MagLab{National High Magnetic Field Laboratory, \protect\\ Tallahassee, FL 32310, USA}

\def\CAS{Beijing National Laboratory for Condensed Matter Physics,
\protect\\Institute of Physics, Chinese Academy of Sciences, Beijing 100190, China}

\def\THU{Beijing National Center for Electron Microscopy and Laboratory of Advanced Materials, \protect\\ Department of Materials Science and Engineering, \protect\\ Tsinghua University, Beijing 100084, China}

\def\ZJU{Center for Correlated Matter and School of Physics, \protect\\ Zhejiang University, Hangzhou 310058, China}

\def\CCQ{Center for Computational Quantum Physics, \protect\\
Flatiron Institute, New York, NY 10010, USA}

\title{Electronic anisotropy and rotational symmetry breaking at a Weyl semimetal/spin ice interface}
\date{}

\author[1, *]{Tsung-Chi Wu}
\author[1,2]{Yueqing Chang}
\author[1,2]{\protect\\ Ang-Kun Wu}
\author[1]{Michael Terilli}
\author[1]{\protect\\ Fangdi Wen}
\author[1]{Mikhail Kareev}
\author[3]{Eun Sang Choi}
\author[3]{\protect\\ David Graf}
\author[4]{Qinghua Zhang}
\author[5]{Lin Gu}
\author[6]{\protect\\ Zhentao Wang}
\author[1,2,7]{Jedediah H. Pixley}
\author[1, *]{Jak Chakhalian}

\affil[1]{\Rutgersphysics}
\affil[2]{\Rutgerscmt}
\affil[3]{\MagLab}
\affil[4]{\CAS}
\affil[5]{\THU}
\affil[6]{\ZJU}
\affil[7]{\CCQ}
\affil[*]{To whom correspondence should be addressed:\protect\\ tcwu@physics.rutgers.edu (T.-C.W.); jak.chakhalian@rutgers.edu (J.C.)}

\begin{document} 

\maketitle 
\baselineskip24pt

\newpage 

\begin{sciabstract}

In magnetic pyrochlore materials, the interplay of spin-orbit coupling, electronic correlations, and geometrical frustration gives rise to exotic quantum phases, including topological semimetals and spin ice. While these phases have been observed in isolation, the interface-driven phenomena emerging from their interaction have never been realized previously. Here, we report on the discovery of interfacial electronic anisotropy and rotational symmetry breaking at a heterostructure consisting of the Weyl semimetal Eu$_2$Ir$_2$O$_7$ and spin ice Dy$_2$Ti$_2$O$_7$. Subjected to magnetic fields, we unveil a six-fold anisotropic transport response that is theoretically accounted by a Kondo-coupled heterointerface, where the spin ice's field-tuned magnetism induces electron scattering in the Weyl semimetal's topological Fermi-arc states. Furthermore, at elevated magnetic fields, we reveal a two-fold anisotropic response indicative of a new symmetry-broken many-body state. This discovery showcases the nascent potential of complex quantum architectures in search of emergent phenomena unreachable in bulk crystals.

\end{sciabstract}

\section*{Introduction}
The interface between two distinct quantum materials offers the rare opportunity to couple seemingly unrelated many-body ground states to create exotic phases that would otherwise be impossible~\cite{zubko2011interface,hwang2012emergent,chakhalian2014colloquium,ramesh2019creating}. 
While there has been a plethora of novel strongly interacting phases discovered in stacked weakly correlated materials, such as graphene- and transition metal dichalcogenide-based heterostructures~\cite{bistritzer2011moire,balents2020superconductivity,andrei2021marvels,yankowitz2019van,kennes2021moire,cao2018unconventional,gibertini2019magnetic,li2021phase,ciarrocchi2022excitonic,huang2020emergent,du2021engineering,mak2018light}, constructing interfaces out of strongly correlated materials, whose individual layers already hold a great deal of intriguing phenomena, is also expected to have significant promise~\cite{zhao2023time, chakhalian2020strongly}.
 
Among the diverse set of strongly correlated materials that show promise for exploring novel interface-driven exotic phases, the magnetic pyrochlore materials (A$_2$B$_2$O$_7$ with A being rare earth and B being Ir or Ti) stand out as particularly hopeful candidates. Here, the many-body phases arise from the interplay of topology, electronic correlations, and magnetic frustration due to their large spin-orbit coupling and unique corner-sharing tetrahedron lattice structure \cite{Gardner2010-tu,rau2019frustrated}. Moreover, the strengths of the electronic correlations and spin-orbit coupling can be tuned by choosing different A and B ions, generating a rich phase diagram of electronic behavior in magnetic pyrochlores \cite{Gardner2010-tu,rau2019frustrated,witczak2014correlated}. It is thus intriguing to devise exotic interfacial quantum states through the thin-film heterostructure assembly of magnetic pyrochlores with inherently different many-body states.

While creating such high-quality interfaces remains a significant experimental challenge, two material classes show promising potential. The first is the pyrochlore iridates, A$_2$Ir$_2$O$_7$ (A = lanthanide ions), that can harbor an antiferromagnetic Weyl semimetal phase with strongly spin-orbit coupled Ir providing itinerant pseudo-relativistic electrons\cite{witczak2014correlated,wan2011topological,chakhalian2020strongly}. The second is the pyrochlore titanates spin ice compounds, X$_2$Ti$_2$O$_7$  (X = Ho and Dy), that are insulating frustrated magnets with emergent electrodynamic excitations governed by strong dipolar interactions~\cite{udagawa2021spin,gingras2010spin,bramwell2001spin,castelnovo2012spin,bramwell2020history,chern2013dipolar}. The synthesis of pristine pyrochlore heterostructures has historically been highly challenging until very recent advances in creating high-quality heterojunction of pyrochlore iridates and titanates using novel hybrid \textit{in-situ} solid state epitaxy method~\cite{kareev2024synthesis}.  Such advancement enables the exploration of exotic interfacial phenomena through the coupling of topological Fermi arcs inherent to the Weyl semimetal with the magnetic excitations characteristic of the spin ice~\cite{chakhalian2020strongly,wen2021epitaxial,liu_magnetic_2021,supplementaryfile}.
Towards this notion, we create a high-quality pyrochlore heterostructure, Eu$_2$Ir$_2$O$_7$/Dy$_2$Ti$_2$O$_7$ (EIO/DTO), which possesses a pristine oriented interface between a Weyl semimetal and classical spin ice. 
The new synthetic structure exhibits a six-fold anisotropic response in angular-resolved magnetotransport upon application of an external magnetic field at ultra-low temperatures. Such surprising behavior is linked to the Kondo-coupled magnetic states of the spin ice DTO, which induce electron scattering in the topological Fermi-arc states of the Weyl semimetal EIO. At higher magnetic fields, the system transitions into a novel rotational symmetry-broken two-fold anisotropic state. This discovery highlights the potential of complex quantum architectures for realizing unique many-body states inaccessible in bulk by transducing exotic magnetic excitations into topologically nontrivial electronic degrees of freedom.

\begin{figure}
    \centering
    \includegraphics[width=\textwidth]{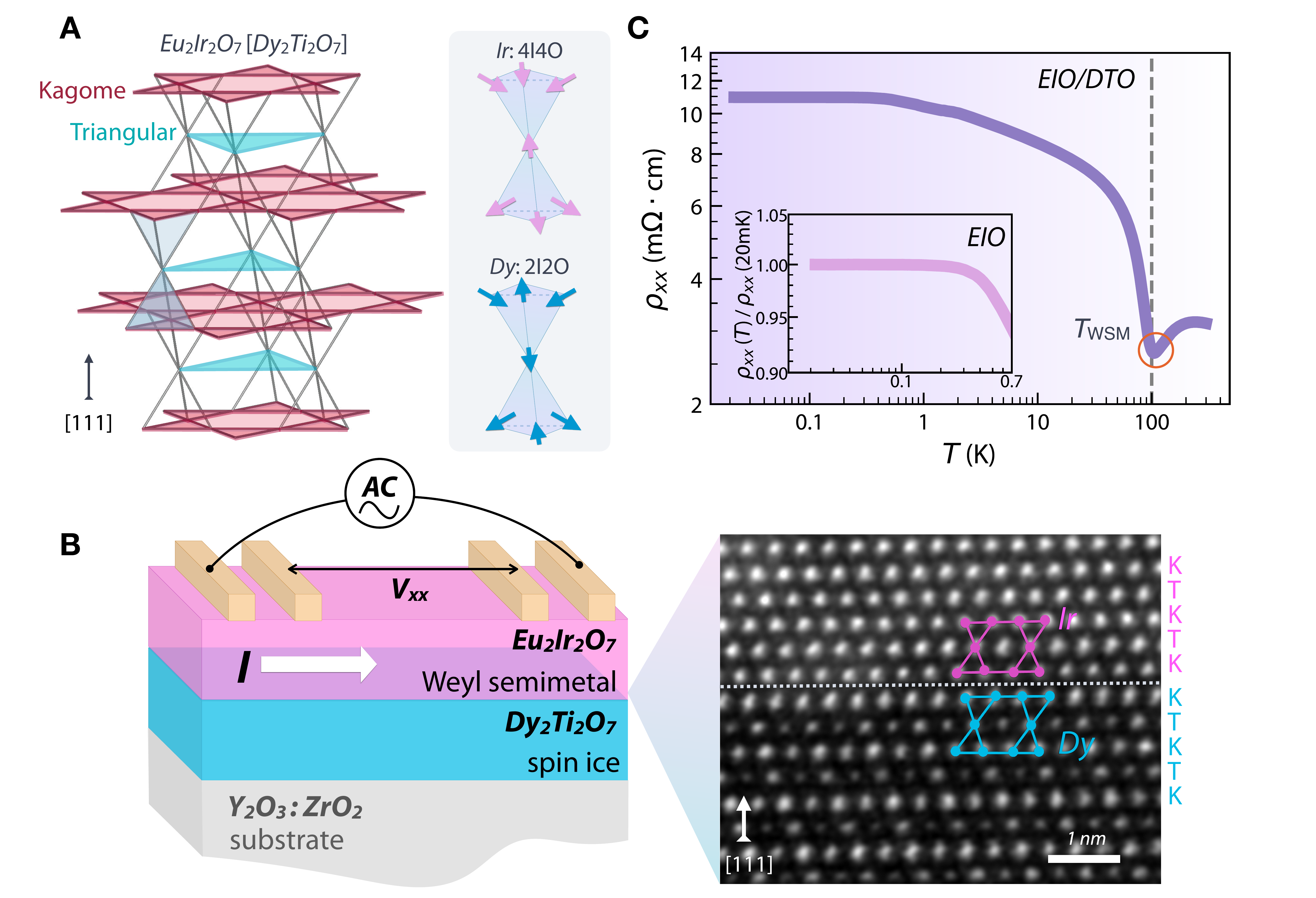}
      \caption{ 
      Crystal structure and temperature dependence of resistivity of EIO/DTO. (\textbf{A}) (Left) Crystal structure of the pyrochlore materials Eu$_2$Ir$_2$O$_7$ (Dy$_2$Ti$_2$O$_7$) viewed along the [111] orientation, whose cations form interpenetrating tetrahedrons. Along the [111] direction,  the sublattices of ions feature an alternating stacking of triangular and kagome planes. Oxygen sublattices are not shown. (Right) The Ir sublattice in Eu$_2$Ir$_2$O$_7$ and the Dy sublattice in Dy$_2$Ti$_2$O$_7$ show four-in-four-out (4I4O) and two-in-two-out (2I2O) spin structures, respectively. (\textbf{B}) Schematic experimental setup for electrical transport measurements on the pyrochlore heterostructure, EIO/DTO, consisting of [111]-oriented thin films of Eu$_2$Ir$_2$O$_7$ (EIO) and Dy$_2$Ti$_2$O$_7$ (DTO) that are Weyl semimetal and spin ice, respectively. EIO/DTO is grown on a non-magnetic insulating substrate, Y$_2$O$_3$:ZrO$_2$ (YSZ). The current is applied along the [1$\bar{\text{1}}$0] direction. The zoom-out view shows the real-space image near the interface using scanning tunneling electron microscopy. The corresponding Ir and Dy ions are shown in pink and blue colors. K and T denote kagome and triangular layers, respectively. (\textbf{C}) Temperature-dependent resistivity $\rho_{xx}(T)$ of EIO/DTO, showing an upturn at $T_{\text{WSM}}$  and a plateau at low $T$. The inset shows the temperature-dependent resistivity data of the control sample, EIO thin films grown on YSZ (denoted as EIO), with $\rho_{xx}(T)/\rho_{xx}$(20$\,$mK), from 20$\,$mK to 700$\,$mK, where a plateau region occurs at $T \leq$ 0.3$\,$K.}
    \label{Fig1}
\end{figure}

\paragraph{Crystal structure and temperature dependence of longitudinal resistivity}\mbox{}\\ 
Both EIO and DTO compounds possess the identical nested corner-sharing tetrahedra of the cations and the global cubic crystal symmetry $Fd\bar{3}m$. As illustrated in Fig. \ref{Fig1}A, when viewed along the [111] direction, each cation sublattice forms an alternating stacking of triangular and kagome planes. Although bulk pyrochlore iridates were one of the first materials candidates predicted to host the magnetic Weyl semimetal phase \cite{wan2011topological,chen2012magnetic,witczak2014correlated}, 
the experimental identification of the phase had been elusive due to the zero anomalous Hall effect (AHE) enforced by the cubic symmetry. Recent theoretical and experimental findings, however,  have demonstrated that films of pyrochlore iridates grown along the [111] direction manifest a pronounced AHE arising from the presence of Weyl nodes,  confirming the topologically non-trivial nature of EIO~\cite{yang2014emergent,liu_magnetic_2021}.

For this study, we synthesized a high-quality heterostructure composed of  EIO and DTO layers along the [111] orientation of the pyrochlore lattice on the insulating substrate 
YSZ (Fig.~\ref{Fig1}B). Notice that the [111] oriented EIO thin film grown directly on the YSZ substrate (EIO) is referred to as the control sample.  At low temperatures, EIO shows the antiferromagnetic Weyl semimetal state with a four-in-four-out (4I4O/4O4I) long-range order of the Ir moments~\cite{wan2011topological,liu_magnetic_2021,liu2024chiral,yang2014emergent}, while DTO exhibits the magnetically degenerate spin ice phase with the imposed two-in-two-out (2I2O/2O2I) spin ice rule on the Dy moments \cite{udagawa2021spin,gingras2010spin,bramwell2001spin,castelnovo2012spin,bramwell2020history} (Fig.~\ref{Fig1}A). Here is an intriguing question: Can passing electrical currents across the EIO/DTO interface transpose the magnetic excitations of the insulating spin ice DTO (Fig.~\ref{Fig1}B) into the electronic degrees of freedom of the conducting Weyl semimetal EIO?

Fig.~\ref{Fig1}C shows the temperature-dependent longitudinal resistivity $\rho_{xx}(T)$
of EIO/DTO with the current ($I$) applied along the [1$\bar{\text{1}}$0] direction.  
 Several characteristic features are immediately seen. First, when cooling down below  300 K, a metal-to-{semimetal}-like transition occurs at $T_{\mathrm{WSM}}$ $\approx$ 105 K, consistent with the previously reported magnetic phase transition of thin-film EIO from a paramagnetic metal phase to an antiferromagnetic Weyl semimetal phase \cite{liu_magnetic_2021}.
Second, upon further decreasing temperatures below 0.3 K, $\rho_{xx}(T)$ develops an unexpected plateau. The measurement on the control sample reveals the presence of both the WSM transition near $T_{\mathrm{WSM}}$ and the plateau region [see fig.~S3 in \cite{supplementaryfile} and inset in Fig.~\ref{Fig1}C]. 
These results establish that the two transport signatures arise from the properties inherent to the EIO layer. Notably, the low-$T$ plateau has been attributed to the presence of topological surface states that are more conductive than the bulk \cite{Resta2018-lu, Breitkreiz2019-ea, Zhang2021-fo, Buccheri2022-ou, Xiang2022-ts, ren2010large, jia2012defects, kim2014topological, dzero2010topological, tafti2016resistivity, Tan2015-uy, Li2020-gh, Liang2018-tc,xu2015discovery,moll2016transport}. For the Weyl semimetal EIO, these topological surface states consist of the Weyl Fermi-arc surface states \cite{wan2011topological,grushin2016inhomogeneous,hasan2017discovery}.

\begin{figure}
    \centering
    \includegraphics[width=\textwidth]{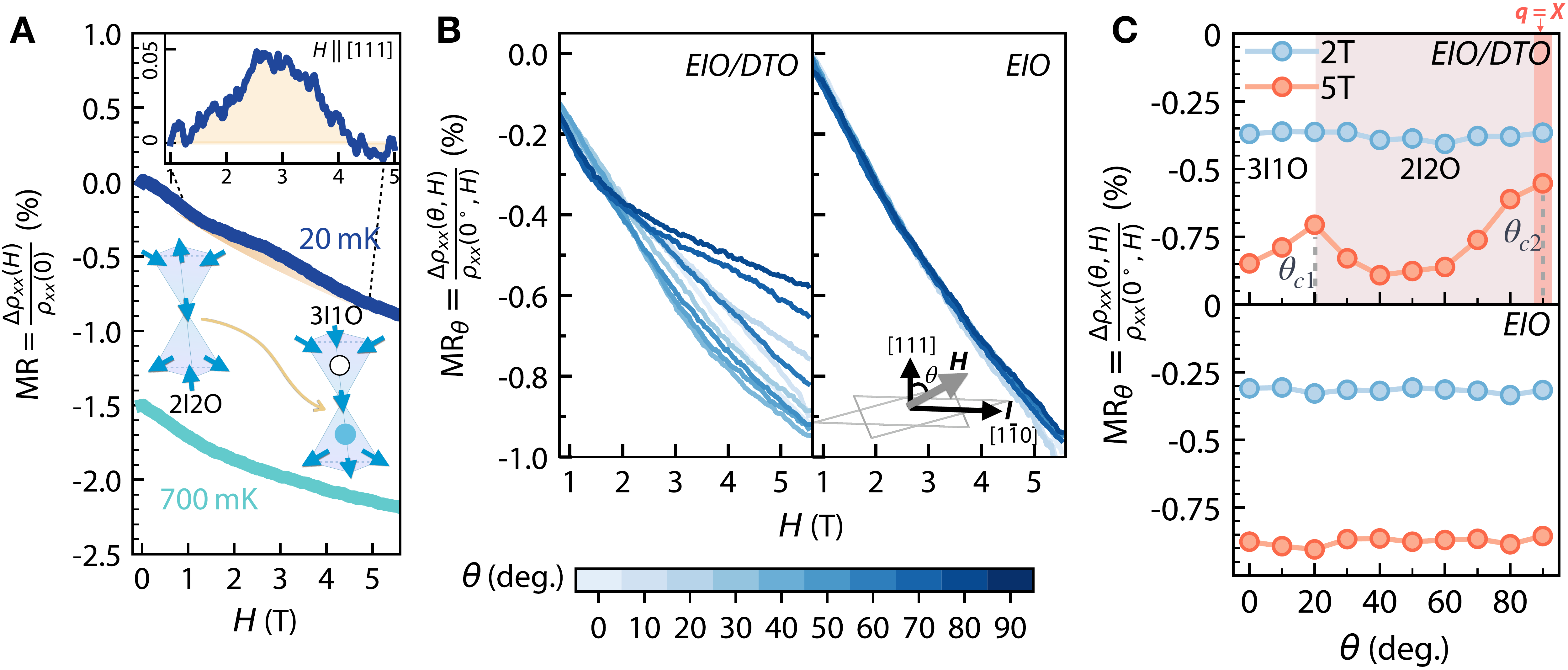}
      \caption{ 
      Angular dependence of magnetoresistance for field rotation away from [111] direction in EIO/DTO. (\textbf{A}) Magnetoresistance (MR) of EIO/DTO with $H$ $\parallel$ [111]. An anomalous MR with a bump feature occurs at 20$\,$mK (blue) (the enclosed yellow area). The inset shows the bump feature after subtracting the background of the raw data [see \cite{supplementaryfile} for the raw data]. In contrast, no bump feature is found at 700$\,$mK (cyan; offset $-1.5\,\%$ for clarity). The left and right tetrahedrons show that the spin structures change from a two-in-two-out (2I2O/2O2I) to a three-in-one-out (3I1O/1I3O) configuration. The latter is the so-called magnetic monopole (MM) state of DTO. The 3I1O and 1I3O spin structures host a monopole (white circle) and anti-monopole (blue circle), respectively. 
      (\textbf{B}) $\text{MR}_{\theta} = \Delta \rho_{xx}(\theta,H)/\rho_{xx}(0^\circ, H)$ of EIO/DTO (left) and the control sample EIO (right) at 20$\,$mK for different field angles, $\theta$. The inset shows the rotational geometry of fields.
      (\textbf{C}) $\text{MR}_{\theta} = \Delta \rho_{xx}(\theta,H)/\rho_{xx}(0^\circ, H)$ of EIO/DTO (top) and the control sample EIO (bottom) at 2$\,$T and 5$\,$T at $\theta=0^{\circ},\,10^{\circ},\,\cdots,\,90^{\circ}$ (extracted from Fig. B). $\theta_{c1}$ and $\theta_{c2}$ are defined as the two peak positions at $20^{\circ}$ and $90^{\circ}$ for $H= 5\,$T, respectively. As $\theta$ increases from $0^{\circ}$ to $\theta_{c1}$ and to $\theta_{c2}$, DTO changes from 3I1O to 2I2O and to $q=X$ phases.
      }
    \label{Fig2}
\end{figure}

\paragraph{Angular dependence of magnetoresistance under \textit{out-of-plane} field rotation}\mbox{} \\
Now, we can investigate the interaction between the Weyl Fermi arcs and spin ice's magnetism. For this purpose, we employ magnetotransport measurements near the resistivity plateau region at sub-Kelvin temperatures. Fig.~\ref{Fig2}A shows the magnetoresistance (MR) of EIO/DTO taken at 20\,mK under an applied magnetic field $H$\,$\parallel$\,[111], where $\textrm{MR}=\Delta\rho_{xx}(H)/\rho_{xx}(0)=(\rho_{xx}(H)-\rho_{xx}(0))/\rho_{xx}(0)$. As immediately seen, a new bump-like feature develops between 1\,T and 5\,T, residing on a background of the negative MR signal characteristic of Weyl semimetals \cite{zyuzin2012topological,xiong2015evidence}. Moreover, the temperature dependence of the magnetoresistance (MR) in the  EIO/DTO heterostructure reveals that the amplitude of the MR feature diminishes with increasing temperatures and vanishes at approximately 700\,mK [see Fig.~\ref{Fig2}A and fig.~S6 in \cite{supplementaryfile}]. In contrast, a careful examination of the control sample shows only the anticipated negative MR signal, devoid of any additional anomalies down to 20\,mK [see fig.~S3 in \cite{supplementaryfile}]. This stark distinction between the EIO/DTO and the singular EIO layer provides compelling evidence that the anomalous transport signature observed in EIO/DTO stems from a non-trivial coupling at the interface, manifesting the presence of a new interface-induced state in EIO/DTO.

 To deepen our understanding of the MR bump-like feature and provide further ground for the interpretation, we recap that in bulk DTO, when a magnetic field is aligned along the [111] direction, it triggers a phase transition that violates the ice rule, transitioning from a  
 kagome ice phase with a two-in-one-out (2I1O/1O2I) configuration within the kagome planes
 to a magnetic monopole phase characterized by three-in-one-out (3I1O/1O3I) spin-dressed tetrahedra shown in Fig.~\ref{Fig2}A \cite{udagawa2021spin,gingras2010spin}.

 By comparing the magnetization ($M$) \textit{vs}. the magnetic field ($H$) between bulk DTO and the heterostructure EIO/DTO configuration, we can draw a direct connection between the observed MR anomaly in the EIO/DTO heterostructure with the ice-rule violating transition into the magnetic monopole phase [for detailed analysis, see Sec. 2.6 in \cite{supplementaryfile}]. 
 This finding supports our conjectures that electric transport under an applied magnetic field acquires sufficient sensitivity to reveal the magnetic phase transition in DTO encoded in the dynamics of Weyl electrons across the EIO/DTO interface. Furthermore, since the magnetic states of DTO can be precisely manipulated through a choice of angles and strengths of the applied magnetic field, we can further probe the nature of the new many-body quantum state in the EIO/DTO heterointerface.

Figure~\ref{Fig2}B shows the angular dependence of the magnetoresistance in EIO/DTO, defined as  $\textrm{MR}_{\theta}=\Delta \rho_{xx}(\theta,H)/\rho_{xx}(0^{\circ},H)=[\rho_{xx}(\theta,H)-\rho_{xx}(0^{\circ},H)]/\rho_{xx}(0^{\circ},H)$, with $H$ rotating away from [111]  to [1$\bar{\text{1}}$0] direction as illustrated in Fig.~\ref{Fig2}B's inset. As seen, above 2\,T we observe a strongly angle-dependent
MR$_{\theta}$ in EIO/DTO, which is entirely absent in the control sample EIO. To further clarify the angular dependence, we plot the values of 
MR$_{\theta}$ at fixed field strengths of 2\,T and 5\,T (Fig.~\ref{Fig2}C). A direct inspection of the results reveals that at 2\,T, no discernible anisotropy is found for both the EIO/DTO and EIO. In contrast, at 5\,T, two distinct peaks develop near $\theta_{c1}=20^{\circ}$ and $\theta_{c2}=90^{\circ}$ only in EIO/DTO.

A straightforward comparison of these observed features in the EIO/DTO with the reported magnetic phase transitions in bulk DTO \cite{udagawa2021spin, gingras2010spin, bramwell2001spin, castelnovo2012spin, bramwell2020history,zhang2023anomalous} confirms that $\theta_{c1}$ aligns with the phase boundary transition from the 3I1O phase to the 2I2O phase while $\theta_{c2}$ marks the transition from the 2I2O phase to an unconventional $q = X$ magnetic phase characterized by antiferromagnetic spin chains oriented perpendicular to the applied field direction [also see fig.~S13 in  \cite{supplementaryfile}]. These findings unambiguously demonstrate the remarkable sensitivity of Weyl fermions at the Fermi arcs of EIO to the spin dynamics of DTO at the EIO/DTO interface, rigorously modulated by the orientation and intensity of the applied magnetic field.

\begin{figure}
    \centering
    \includegraphics[width=\textwidth]{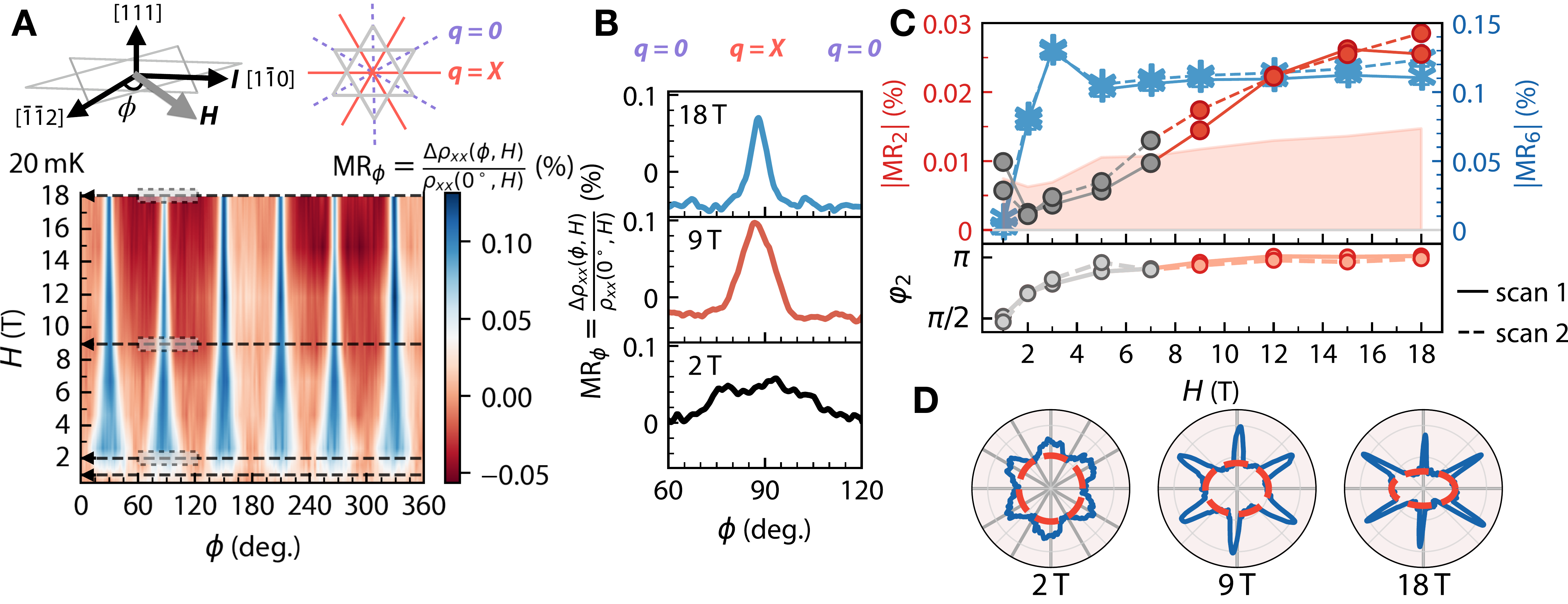}
      \caption{ 
      Angular dependence of magnetoresistance for field rotation within (111) plane in EIO/DTO. (\textbf{A}) (Top) The left scheme shows the geometry for field rotation angles, $\phi$.  The right scheme shows that $q = 0$ (dash line) and $q=X$ (solid line) phases of DTO happen at $\phi = 60n^\circ$ and $\phi = (60n+30)^\circ$, respectively, where n $=$ 0, 1, 2, 3, 4, and 5.
      (Bottom) Contour plot of $\text{MR}_{\phi} = \Delta \rho_{xx}(\phi,H)/ \rho_{xx}(0^\circ,H)$ at 20$\,$mK in EIO/DTO for $H$ up to 18 T. For $H = 2\,$T (black dash arrow), six broad peaks (blue regions) and uniform backgrounds (white-red regions) are found, revealing a six-fold symmetry. For $H = 9, 18\,$T (black dash arrow), a dark red background starts to develop, showing a color contrast in the backgrounds where the six-fold symmetry is broken into the two-fold symmetry. 
      (\textbf{B}) $\text{MR}_{\phi} = \Delta \rho_{xx}(\phi,H)/\rho_{xx}(0^\circ,H)$ at 20$\,$mK for $\phi$ from $60^\circ$ to $120^\circ$ at 2$\,$T (black), 9$\,$T (red), and 18$\,$T (blue) (see the grey rectangles in panel A). The narrowing of the MR peak at $90^\circ$ is found while increasing magnetic fields. While the low MR values happen at $60^\circ$ to $120^\circ$ where DTO is in $q = 0$ state, the high MR value happens at $90^\circ$ where DTO is in the $q = X$ state.  
      (\textbf{C}) Magnitudes of the two-fold and six-fold contributions to $\text{MR}_{\phi}$, i.e., $\text{MR}_2$ and $\text{MR}_6$, respectively [see the main text and Sec.~2.9 of \cite{supplementaryfile} for details], plotted in red circles and blue stars versus the magnetic field strengths, for the two datasets, i.e., scan 1 and scan 2 [see fig.~S2 in \cite{supplementaryfile}].
      The shaded red region shows the estimated fitting uncertainty obtained using Bayesian inference [see Sec.~2.9 of \cite{supplementaryfile}].
      From 2\,T to 7\,T, the magnitude of the two-fold contribution $|\text{MR}_2|$ lies below the uncertainty of the fitting.
      For higher magnetic fields above 9\,T, the two-fold anisotropy is clearly visible 
      and 
      shifts in its phase by $\pi/2$, as shown by the fitted $\varphi_2$ in (C) (also see (D)).
      On the other hand, the six-fold contribution $|\text{MR}_6|$ is present from 2$\,$T to 18$\,$T.
      (\textbf{D}) Polar plots of $\text{MR}_{\phi}$ (of scan 2) at 20$\,$mK at 2, 9, and 18$\,$T (indicated by the black dashed line cuts in Fig.~\ref{Fig3}A), where the solid blue lines, solid grey lines, and red dashed lines denote the $\text{MR}_{\phi}$ data, mirror planes, and oval backgrounds, respectively. 
      $\text{MR}_{\phi}$ shows a 
      a six-fold anisotropy at 2$\,$T and a two-fold anisotropy at 9 and 18$\,$T (see the oval backgrounds).
      }   
    \label{Fig3}
\end{figure}

\begin{figure}
    \centering
    \includegraphics[width=\textwidth]{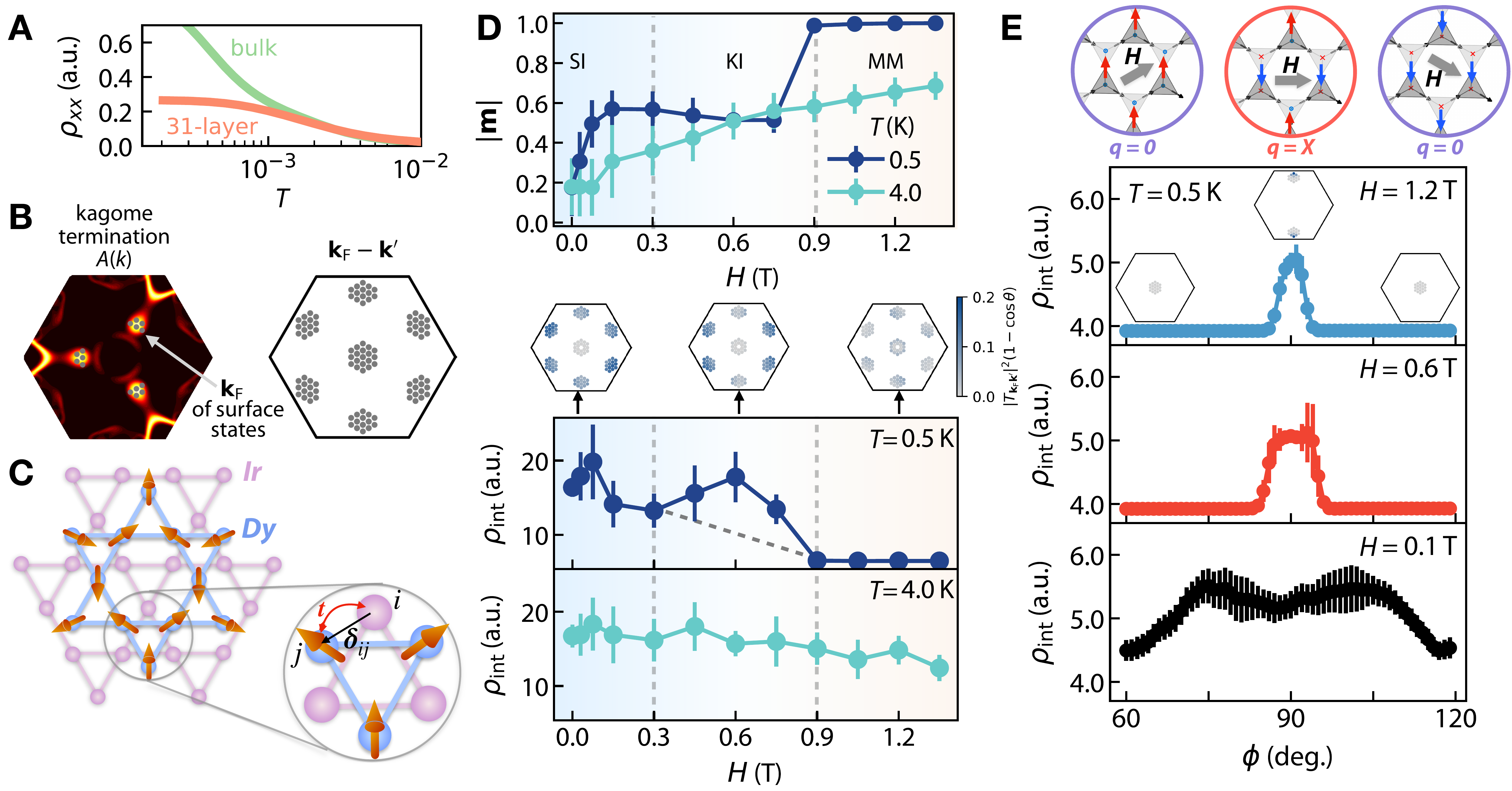}
    \caption{
    Phenomenological theoretical description
    of the observed transport behavior in  
    EIO/DTO. 
    (\textbf{A})
    Theoretically computed $\rho_{xx}$ versus temperature $T$ in bulk EIO and 31-layer EIO slab.
    $\rho_{xx}$ shows a saturated plateau at low-$T$ in the slab but not in the bulk, demonstrating the effect of metallic surface states present in the slab.
    (\textbf{B})
    The Fermi arcs of the EIO slab at the kagome termination, where the Fermi vectors $\mathbf{k}_{\text{F}}$, corresponding to the three small electron pockets at the projected Weyl points, are highlighted by the grey dots.
    On the right, we show the momenta where the integration of $\tau^{-1}_{\text{int}} (\mathbf{k}_{\text{F}})$ is performed: $\mathbf{q} =  \mathbf{k}_{\text{F}} - \mathbf{k}^\prime$, where $\mathbf{k}'\in \left\lbrace \mathbf{k}_{\text{F}}\right\rbrace$.
    (\textbf{C})
    The interface of EIO and DTO realizes a Kondo coupling between the kagome Ir-lattice and kagome Dy-lattice. 
    To simplify, we include up to the nearest-neighbor Kondo coupling between the Ir and Dy orbitals $i$ and $j$ separated by vector $\delta_{ij}$ and assume that the coupling strength $J_{ij}\propto |t|^2$ is orbital independent. 
    (\textbf{D}) When applied with an increasing magnetic field along the [111] direction, the DTO first shows a cross-over from the spin ice (SI) phase to the kagome ice (KI) phase, followed by an ice-rule-breaking transition to the magnetic monopole (MM) phase (upper panel). 
    The computed interfacial resistivity $\rho_{\text{int}}$ shows a bump-like feature (lower panel), consistent with the experiments (see Fig.~\ref{Fig2}A).
    The insets between two panels show the integral kernel of $\tau_{\text{int}}^{-1}(\mathbf{k}_{\text{F}})$, i.e., 
    $|T_{\mathbf{k_{\text{F}}k}'}|^2(1-\cos(\theta))$, where darker colors reveal larger contributions to the integral.
    The critical fields by the Monte Carlo calculation are consistent with the reported bulk calculations, and their deviation from the experimental values is discussed in Sec.~2.8.2 of \cite{supplementaryfile}.
    (\textbf{E}) 
    The computed $\rho_{\text{int}}$ versus $\phi$, for $60^\circ \leq \phi \leq 120^{\circ}$ under in-plane magnetic fields $H$ perpendicular to [111].
    At $\phi=60^\circ, 120^\circ$, $\rho_{\text{int}}$ is small while DTO shows a zero-momentum phase ($q=0$).
    At $\phi=90^\circ$, $\rho_{\text{int}}$ is large while DTO shows a finite-momentum phase ($q=X$) as the field partially unpins the magnetic frustrations in the kagome Dy-lattice. 
    As the field strength increases, the peaks in $\rho_{\text{int}}$ become narrower, in agreement with the observed angular narrowing of $\text{MR}_{\phi}$ at high fields in experiments (see Fig.~\ref{Fig3}B).
}
    \label{Fig4}
\end{figure}

\paragraph{Angular dependence of magnetoresistance under \textit{in-plane} field rotation}\mbox{}\\
Having established the strong link between the Weyl fermions and specific spin ice magnetic configurations in EIO/DTO, we can now explore the microscopic characters of the interfacial states arising from such coupling. 
For this purpose, we turn to the angular-dependent MR measurement under in-plane applied magnetic fields at an angle $\phi$, MR$_{\phi}$ (see Fig.~\ref{Fig3}A for the definition and geometry).
 A direct inspection of the data shown in Fig.~\ref{Fig3}B unveils several remarkable features. First, we identify a surprising six-fold rotational symmetric state that emerges after the magnetic field exceeds 2\,T. Second, the transport signal shows a distinctive continuous angular narrowing at the specific $\phi_{n} = (30^{\circ}+60^{\circ}n)$  ($n=0...5$) directions where the magnetoresistance reached its maxima. The lowest values of MR$_{\phi}$ are found in six $\phi_n$ directions along $60^{\circ}n$. To understand the six-fold anisotropic state at the interface between the EIO and DTO layers, we refer to Fig.~\ref{Fig3}A and \ref{Fig4}E. Previous extensive investigations on single crystal DTO \cite{Higashinaka2005-vj, Sato2006-to, Matsuhira2007-gr, Kao2016-yx} revealed that an in-plane magnetic field effectively controls the magnetic state of DTO by switching its magnetism from  $q = 0$ phase at $\phi_n = (60^{\circ}n)$ to $q = X$ phase at $\phi_n = (30^{\circ}+60^{\circ}n)$. Remarkably, the appearance of the six-fold symmetric MR$_{\phi}$ is in excellent registry with the magnetic field-tuned cyclical switching between the $q = 0$ and $q = X$  spin structures of DTO. Further, to shed light on the unusual angular narrowing, we plot MR$_{\phi}$ at 2, 9, and 18\,T for $\phi$ varying from $60^{\circ}$ to $120^{\circ}$ (see Fig.~\ref{Fig3}B and the grayed area in Fig.~\ref{Fig3}A). 
By analyzing angular MR$_{\phi=90^{\circ}}$  which scales with  $\Delta\rho_{xx}(H)/\Delta\phi$, we reveal a gigantic almost 600$\,\%$ increase in the angular magnetoresistance value  at 18\,T compared to that of 2\,T [see Fig.~S7 in \cite{supplementaryfile}].  Notably, the angular narrowing phenomenon is rare, closely resembling the large angular response reported in magnetic nodal crystals \cite{suzuki2019singular,seo2021colossal}.

In addition to these two observations, another striking finding is the emergence of a novel quantum state when the applied magnetic field exceeds 9 T, as illustrated in Fig.~\ref{Fig3}B. At lower magnetic fields, just above 2\,T, the magnetoresistance $\text{MR}_{\phi}$ exhibits six broad peaks against a uniform backdrop, implying a six-fold rotational symmetry. This symmetry transforms drastically once the magnetic field exceeds 9\,T, where the $\text{MR}_{\phi}$ signal traverses into a two-fold or bilateral symmetry state. Polar plots of the magnetoresistance taken at 2, 9, and 18\,T  shown in Fig.~\ref{Fig3}C  demonstrate this transition. At 2\,T, the polar plot displays a six-fold rotational symmetry with an almost perfect circular background, consistent with the symmetry expected from the field-tuned $q = 0$ and $q = X$ magnetic states. However, beyond 9\,T, a distinct transition to a two-fold symmetry with an oval-like backdrop is observed. Interestingly, the degree of ovality also increases progressively up to 18\,T [see  fig.~S18 in \cite{supplementaryfile}].

A comprehensive analysis of the MR signal [see Sec.~2.9 of \cite{supplementaryfile}]  reveals the evolution of two distinct contributions to the magnetoresistance, namely, the six-fold $|\text{MR}_{6}|$ and the two-fold $|\text{MR}_{2}|$ as a function of the magnetic field. While the $|\text{MR}_{6}|$ signal remains finite and nearly unchanged from 2\,T to 18\,T, the $|\text{MR}_{2}|$ signal appears right above 9\,T and grows with increasing magnetic field strength up to 18\,T. The absence of both field-induced features in MR$_{\phi}$ for the control sample emphasizes that the unique six-fold and two-fold anisotropic states observed at $H\geq2$ T and $H\geq9$ T, respectively, are exclusively induced by the interface-coupled Weyl electrons of EIO and magnetic excitations of DTO [see fig.~S3 in \cite{supplementaryfile}]. The new two-fold quantum state with strong electronic anisotropy and broken rotational symmetry represents an emergent phase unattainable in bulk.

\paragraph{Modeling of the Weyl semimetal/spin ice interface}\mbox{}\\
In what follows, we describe a microscopic framework for understanding the observed magnetotransport phenomena in EIO/DTO, providing a qualitative explanation for the experimental findings [see Sec.~2.8 in \cite{supplementaryfile} for more details]. To achieve this, we devise an effective model for the heterostructure's Hamiltonian, 
$\hat{\mathcal{H}}_{\text{EIO/DTO}} = \hat{\mathcal{H}}_{\text{EIO}} + \mathcal{H}_{\text{DTO}}+\hat{\mathcal{H}}_{\text{int}}$, where the last term describes the interfacial interaction between the Dy and Eu states placed on the kagome lattice. We solve for $\hat{\mathcal{H}}_{\text{EIO}}$ and $\mathcal{H}_{\text{DTO}}$ separately before coupling them through a perturbative treatment of $\hat{\mathcal{H}}_{\text{int}}$.   First, we establish that  EIO's Weyl Fermi arcs expressed in $\hat{\mathcal{H}}_{\text{EIO}}$ can give rise to a resistivity plateau at low temperatures. As shown in Fig.~\ref{Fig4}A, the computed $\rho_{xx}(T)$ for EIO clearly reproduces the experimentally found low-temperature plateau in longitudinal resistance $\rho_{xx}(T)$ which occurs only in the EIO slab but not in bulk EIO, thus attributing the plateau to the presence of the Weyl Fermi-arc surface states shown
in the spectral function $A(\mathbf{k})$ projected on a kagome terminated atomic plane in Fig.~\ref{Fig4}B.

To capture the physics encoded in the $\mathcal{H}_{\text{DTO}}$ term, we use a dipolar classical spin model solved with a classical Monte Carlo (MC) approach at low temperatures. Based on those approaches, we accurately reproduce the reported $M$-$H$ curve for bulk kagome ice and magnetic monopole phases under a [111] applied magnetic field shown in Fig.~\ref{Fig4}D, upper panel [also see Sec.~2.8.2 and fig.~S6 in \cite{supplementaryfile}].

Next, we consider the interfacial interaction dominated by the Kondo coupling between the Fermi-arc surface states and the local Dy moments. As pictured in Fig.~\ref{Fig4}C, this coupling is represented by $\hat{\mathcal{H}}_{\text{int}}=\sum_{\langle i,j \rangle}J_{ij}{\bf S}_i\cdot{\bf \hat{s}}_j$, where $J_{ij}$ is the interfacial superexchange between Dy (classical moments ${\bf S}_i$) and Ir (quantum moments $\hat{{\bf s}}_j$), i.e. Kondo coupling, and the sum is over the nearest neighbors between the two interfacial layers.  Within the Boltzmann approximation \cite{wang_resistivity_2016}, the coupling term introduces an interfacial resistance, $\rho_{\text{int}}$, by inducing a finite relaxation time for the conducting Fermi-arc surface states, i.e., 
$\rho_{\text{int}}  \propto$ 
$\tau^{-1}({\bf k}_F) \propto \int \mathrm{d}{\bf k^{\prime}} 
|T_{\mathbf{k}_{\text{F}}{\mathbf{k}'}}|^2
    \delta(\epsilon_{\text{F}} - \epsilon_{\mathbf{k}'})
 (1-\cos\theta_{\mathbf{k}_{\text{F}}\mathbf{k}'})$, where $\epsilon_{\mathbf{k}}$ is the dispersion of the surface states of EIO, and $\theta_{\mathbf{k}_{\text{F}}\mathbf{k}'}$ is the scattering angle. The scattering matrix element $|T_{\mathbf{k}_{\text{F}}{\mathbf{k}'}}|^2$ is proportional to the DTO's spin structure factor $S({\bf q}) = \frac{1}{N}\langle {\bf S}_{\bf q}\cdot {\bf S}_{-{\bf q}} \rangle$ 
and the set of $\bf q = \bf k_{\text{F}} - \bf k'$ that dominates this interfacial response is highlighted in Fig.~\ref{Fig4}B. Intuitively, the overlap between $A(\mathbf{k})$ of EIO and $S({\bf q})$ of DTO are the prime factors that dictate the magnitude of the interfacial resistance $\rho_{\text{int}}$.

Within this model, we compute $\rho_{\text{int}}$ versus $H$ applied along  [111], which reveals an anomalous bump feature in good agreement with the experimental observation (Fig.~\ref{Fig4}D and \ref{Fig2}A). Interestingly, the calculation asserts that the bump feature develops when the DTO layer enters the kagome ice phase and disappears in the magnetic monopole phase. In particular, the integral kernel, $|T_{\mathbf{k}_{\text{F}}\mathbf{k}'}|^2(1-\cos\theta_{\mathbf{k}_{\text{F}}\mathbf{k}'})$ (Fig.~\ref{Fig4}D inset), demonstrates that the decrease in $\rho_{\text{int}}$ arises from the vanishing of $S(\mathbf{q})$ near the Brillouin zone (BZ) boundary upon increasing magnetic fields.

Further, with a magnetic field $H$ aligned perpendicular to [111], the computed interfacial resistivity, $\rho_{\text{int}}$, accurately yields the experimentally observed six-fold anisotropic behavior and the angular narrowing of MR$_{\phi}$ (see Fig.~\ref{Fig3}). As shown in Fig.~\ref{Fig4}E, the extrema in $\rho_{\text{int}}$ coincide with the $q=X$ and $q=0$ phases of DTO, achieving maximum and minimum MR$_{\phi}$ values, respectively. More precisely, because of the negligible lattice mismatch at the interface, $\rho_{\text{int}}$ reaches its maximum as the integral kernel $|T_{\mathbf{k}_{\text{F}}\mathbf{k}'}|^2(1-\cos\theta_{\mathbf{k}_{\text{F}}\mathbf{k}'})$ contributes most near the BZ boundary (Fig.~\ref{Fig4}E, top panel insets). Here, the scattering vectors $\mathbf{q}$ from the Weyl pockets of EIO strongly `resonate'  with DTO's structure factor $S(\mathbf{q})$, thus preferentially picking up the signal in the $q=X$ phase compared to the $q=0$ phase. In addition, as magnetic field strength keeps increasing, the $q=X$ phase in DTO appears within a narrow angular range around $90^\circ$ of $\phi$. As such, the resultant $\rho_{\text{int}}$ peaks become narrower at higher magnetic fields, accounting for the experimental angular narrowing effect in MR$_{\phi}$ (see Fig.~\ref{Fig4}E).

Overall, our theoretical framework provides a coherent explanation for the observed highly anisotropic transport phenomena in angular-dependent magnetoresistance, notably the six-fold anisotropic behavior and angular narrowing effect. The observed responses can thus be attributed to the magnetic-field-tuned interplay between EIO's Weyl Fermi arcs, the scattering factor of DTO's magnetic states, and the interfacial Kondo coupling between the Weyl fermions and Dy spins. Finally, the inability of $\hat{\mathcal{H}}_{\text{int}}$ to capture the two-fold symmetry-broken state emerging at 9\,T implies the non-perturbative and conceivably non-local nature of the state, which calls for further exploration.


\paragraph{Discussion and Outlook}\mbox{}\\
It is a striking observation that the topological surface states of EIO exhibit the sensitivity required to detect the distinct magnetic phases of DTO. This detection is feasible only when the geometry of the topological Fermi arcs has sufficient overlap with the magnetic structure factor of DTO. Also, our calculations emphasize the crucial importance of the Weyl semimetal phase of EIO  for the anisotropic transport behavior at the EIO/DTO  interface. Remarkably, due to the intertwined nature of this coupling, the anisotropic transport response in EIO/DTO provides unique microscopic insights into not only the magnetic properties of DTO but also the topological properties of EIO.

The ability to explore the complex magnetic states of spin ice via the electrical properties in the EIO/DTO heterostructure is central to our findings of the multifaceted interface-driven phenomena, including the magnetic-field-induced electronic anisotropy and rotational symmetry breaking. The demonstrated interfacial phenomena within the EIO/DTO pyrochlore heterostructure underpin the strategy that can be extended to seek proximity-induced topology and exotic magnetism across the spectrum of frustrated quantum pyrochlores where conventional probes fail \cite{rau2019frustrated}. For instance, in EIO/DTO, replacing Dy with Tb within the DTO layer transforms it from classical into quantum spin ice \cite{rau2019frustrated, Shannon2012-zl, Ross2011-oi, Gardner1999-mk, Gardner2001-fd, bojesen2017quantum,hallas2018experimental}, creating 
an unprecedented synthetic two-dimensional heavy-fermion structure in the presence of emergent dynamical gauge fields. Finally, our discovery of the rotational symmetry-broken electronic states at the interface validates the feasibility of magnetic pyrochlore architectures to realize new quantum phases inaccessible in bulk crystals.

\newpage

\section*{References and Notes}

\nocite{PhysRevLett.107.196803}
\nocite{PhysRevLett.107.127205}
\nocite{RevModPhys.78.275}
\nocite{PhysRevLett.114.116602}
\nocite{joao2019basis}
\nocite{wu2023absence}
\nocite{denHertog_dipolar_2000}
\nocite{mahan2013many}
\nocite{pymc2023}

\bibliographystyle{science}





\newpage

\section*{Acknowledgments}
We acknowledge valuable discussions with D. Vanderbilt, P. Chandra, P. Coleman, D. Kaplan, S. Fang, V. Drouin-Touchette, P. Orth, W.-H. Kao, and Y.-J. Kao. 
We also acknowledge the technical support from R. Nowell. 
\textbf{Funding:}
J.C., T.-C.W., M.T., and M.K. acknowledge the support by the U.S. Department of Energy, Office of Science, Office of Basic Energy Sciences under Award No. DE-SC0022160.  
This research is funded in part by a QuantEmX grant from ICAM and the Gordon and Betty Moore Foundation through Grant GBMF9616 to T.-C.W.
A portion of this work was performed at the National High Magnetic Field Laboratory, which is supported by National Science Foundation Cooperative Agreement No. DMR-1644779 and DMR-2128556 and the State of Florida. 
Q.Z. and L.G. acknowledge the support of the National Natural Science Foundation of China (52250402, 52025025).
Y.C. acknowledges the support of the Abrahams Postdoctoral Fellowship from the Center for Materials Theory, Department of Physics and Astronomy at Rutgers University.
A.-K.W. and J.H.P. acknowledge NSF Career Grant No. DMR-1941569 and the Alfred P. Sloan Foundation through a Sloan Research Fellowship.
Z.W. acknowledges the support of the National Natural Science Foundation of China Grant No. 12374124.
\textbf{Competing interests:} The authors declare no competing interests.
\textbf{Data and materials availability:} 
All data are available in the manuscript or the supplementary materials.

\newpage

\section*{Supplementary materials}
Materials and Methods\\
Supplementary Text\\
Figures S1 to S18\\
References \textit{(69-77)}

\clearpage

 \section*{Figures}
\clearpage

\end{document}